\documentclass[epj,final]{svjour}
\usepackage[utf8x]{inputenc}
\usepackage{amsmath,amssymb,graphicx}
\newcommand{\avg}[1]{\langle #1\rangle}
\newcommand{\onefig}[1]{\includegraphics[scale=0.31]{#1}}
\begin{document}
\title{Self-organized model of cascade spreading}
\author{Stanislao Gualdi, Matúš Medo, Yi-Cheng Zhang}
\institute{Physics Department, University of Fribourg,
CH-1700 Fribourg, Switzerland}
\abstract{We study simultaneous price drops of real stocks and
show that for high drop thresholds they follow a power-law
distribution. To reproduce these collective downturns, we
propose a~minimal self-organized model of cascade spreading
based on a probabilistic response of the system elements to
stress conditions. This model is solvable using the theory of
branching processes and the mean-field approximation. For a wide
range of parameters, the system is in a critical state and
displays a~power-law cascade-size distribution similar to the
empirically observed one. We further generalize the model to
reproduce volatility clustering and other observed properties of
real stocks.}

\maketitle

\section{Introduction}
Cascade spreading is an important emergent property of various
complex systems. Real life examples of cascades are numerous and
range from infrastructure failures and epidemics to traffic jams
and cultural fads~\cite{Glad00,DCLN07}. Theoretical models of
cascades usually assume that agents can be in one of two states
(healthy or failed) and an agent's failure puts some stress on
its neighbors which may consequently fail too.
See~\cite{LoBaSch09} for a recent survey of this field offering
a novel unifying view.

In this paper we focus on cascades in economic systems which
can be identified with stock prices suddenly dropping in a~major
market crash~\cite{Sorn03} or with companies going bankrupt
simultaneously and leading to global recession~\cite{HoLeLe07}.
Theoretical models of such cascades are based on shortage and
bankruptcy propagation in production networks~\cite{WeBa07},
default propagation in credit networks~\cite{IoJa01,SiHo09},
interaction of firms through one monopolistic bank~\cite{IAFIS09}
or in a~complex credit network economy~\cite{DGGRS09}, and
herding behavior of traders~\cite{BaPaSh97,CoBo00}. While these
models help us to understand cascade processes in economic
systems, they are mostly too involved to allow for analytical
solutions---their study hence relies on numeric simulations and
agent-based modeling~\cite{MiPa07}.

A simpler point of view on cascade phenomena is offered by the
concept of self-organized criticality (SOC) which has had a~deep
impact on the science of complexity. First introduced more than
twenty years ago to explain the ubiquitous $1/f$
noise~\cite{BTW87}, it caused a blossoming of toy models,
computer simulations, and real life experiments~\cite{Turc99}.
The analytical techniques employed include scaling
arguments~\cite{TaBa88}, mean-field theories~\cite{FlySneBa93},
branching processes~\cite{Als88}, renormalization
methods~\cite{PiVeZa94,Mars94}, and rigorous algebraical
techniques~\cite{Dhar90}.

SOC is a mechanism which explains the emergence of complex
behavior in many diverse real world systems~\cite{PaBa99,Sorn06}.
The generic behavior of SOC models is: (a) they evolve so that
they always stay close to the critical point, (b) long periods
of robustness and moderate activity are interrupted by sudden
breakdowns. This qualitatively resembles ``stock markets which
expand and grow on relatively long time scales but contract in
stock-market crashes on relatively short time
scales''~\cite{Turc99} and ``stock crashes caused by the slow
buildup of long-range correlation leading to a global
cooperative behavior of the market eventually ending into
a~collapse in a short time interval''~\cite{Sorn03}. This
similarity provides the main motivation for the present study.

We begin our work with an empirical investigation of
simultaneous price drops of real stocks and show that the size
distribution of observed events is broad (for high drop
thresholds it follows a power-law distribution). This
observation suggests that simultaneous stock downturns are a
collective phenomenon. We propose a simple dynamical model which
for a wide range of parameters self-organizes into a critical
state. Unlike most SOC models, our model assumes a probabilistic
response mechanism where a node has only a certain probability
of reacting to the current stress conditions. The basic idea
behind modeling simultaneous stock downturns with cascades is
that decline of a single stock may provoke investors' reactions
which consequently may cause other stocks to decline and a
``cascade'' to spread. The key premise is that while failed
nodes become significantly more resistant in the next time step,
healthy nodes become slightly less resistant. This close
parallel with the slow growth/fast decay picture described above
is further supported by our analysis of empirical data which
shows that majority of stocks behave in this way. While there
are certainly many other effects contributing to the dynamics of
market crashes (external shocks, for example), we show that
failure propagation alone can reproduce some of the observed
patterns.

The minimal model proposed here has the advantage of being
simple, not relying on fine-tuning of parameters, analytically
solvable in some cases, and easily generalizable to more
complicated settings. We analyze it using the formalism of
branching processes, the mean-field approximation and, for
complex topologies of nodes' interactions, using numerical
simulations. Obtained cascade-size distributions exhibit a close
similarity to our empirical observations. Introduction of memory
within the model allows us to reproduce other empirically
observed features, such as volatility clustering, though at the
cost of analytical tractability. We conclude our study with a
discussion of further model's generalizations and possible areas
of application.

\section{Empirical data}
\label{sec:empirical}
Here we investigate co-occurring price movements of real stocks.
Adopting the vocabulary of cascade models, we say that a stock
fails when the relative loss of its price over a given time
interval $\Delta t$ exceeds a certain threshold $H$. Denoting
the price of stock $i$ at time $t$ as $p_i(t)$, its failure
occurs when $[p_i(t)-p_i(t+\Delta t)]/p_i(t)>H$. The number of
stocks failing at time $t$, $n_F(t)$, is a direct analog of the
cascade size in a model of cascade spreading. As the input data
we use daily closing prices (hence $\Delta t=1\,\mathrm{day}$)
of 500 stocks from the standard U.S. index S\&P 500 (this data
is freely available at, for example, \texttt{finance.yahoo.com}).
To achieve a fixed system size, we consider only those 332
companies which are in the stock market since the beginning of
1992 and use their prices during the 18-years long period ending
in May 2010 for our analysis.

\begin{figure}
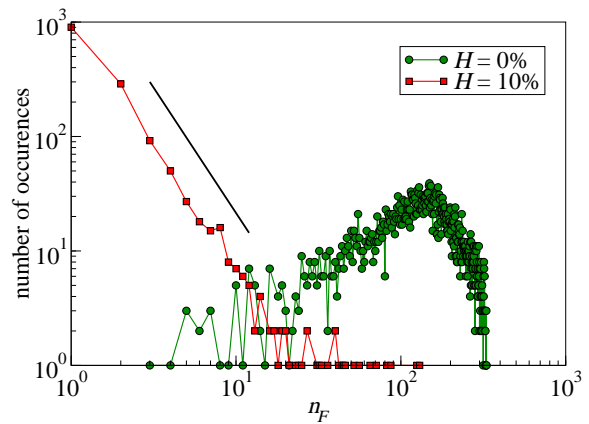

\centering
\vspace*{4pt}
\onefig{P_S-real}
\caption{The empirical failure size distribution observed with
real stock prices (daily closing prices of 332 companies from
January 1992 until May 2010) for threshold relative drops
$H=0\%$ and $H=10\%$. The straight line corresponds to the
exponent $2.19$ obtained by statistical analysis of the data.}
\label{fig:real_stocks}
\end{figure}

The empirical distribution of failure sizes is shown in
Fig.~\ref{fig:real_stocks} for $H=0\%$ and $H=10\%$. We see that
for the large value of $H$ (which is in line with the notion of
stock \emph{failures}), the observed size distribution has a
power-law shape. Using the methodology described
in~\cite{ClaShaNew09}, we obtained the power-law exponent
$2.19\pm0.05$ with the lower bound for the power-law behavior
$n_{\min}=3$. The corresponding $p$-value (obtained using the
standard Kolmo\-go\-rov-Smirnov statistic) is 0.92 which confirms
that the data is consistent with the hypothesis of a power-law
distribution. Similar results are obtained also for other
threshold values so long as $H\gtrsim8\%$. When $H\lesssim8\%$,
the resulting size distributions are broad but probably not
power-law. Finally, when $H=0\%$ (\emph{i.e.}, any price drop is
interpreted as a failure), the size distribution is roughly
symmetric around the value corresponding to one half of the
system size (see Fig.~\ref{fig:real_stocks}). In the following
analysis of empirical data we use the threshold $H=10\%$.

The power-law shape itself suggests that the observed
simultaneous stock downturns are rather a collective phenomenon
than independent events. This hypothesis is further supported by
the average correlation of simultaneously failing stocks, $0.35$
(again including only events with at least three simultaneously
failing stocks), which is significantly higher than the overall
average stock correlation, $0.25$. Another sign of a strong
connection among simultaneously failing stocks comes from their
division to ten different industrial sectors according to the
GICS classification. The effective number of sectors
participating in a cascade is defined as
\begin{equation}
e:=\bigg(\sum_{i=1}^{10} r_i^2\bigg)^{-1}
\end{equation}
where $r_i$ is the relative share of sector $i$ in the cascade
and $\sum_{i=1}^{10} r_i=1$. By averaging this quantity over all
cascades of a given size $S$, we obtain $e(S)$. This number can
be compared with the effective number of sectors corresponding
to selecting failed stocks at random, $e'(S)$. The analysis of
stock prices shows that for any $S>3$, $e(S)$ is significantly
smaller than $e'(S)$ which implies that simultaneous stock
failures preferentially affect strongly connected stocks in one
sector or in a small number of sectors.

Now we turn our attention to time correlations of failures. The
autocorrelation of the number of failing stocks with the time
lag one day, $C(n_F(t),n_F(t+1))\approx 0.15$, is comparable
with the autocorrelation of absolute returns,
$C(\vert r(t)\vert,\vert r(t+1)\vert)\approx0.25$
(the latter result agrees with previous
studies~\cite{Akg89,ManSta99}). The positive autocorrelation
values are signs of volatility clustering which is commonly
observed in financial data~\cite{Cont07}. (Loosely speaking,
volatility clustering means that large changes tend to be
followed by large changes and small changes tend to be followed
by small changes, as first noted by Mandelbrot~\cite{Mand63}.)

We further estimate conditional failure probabilities for
individual stocks. For example, $P(F\vert N)$ denotes failure
probability of a stock given that this stock didn't fail in the
previous time step (other three quantities, $P(N\vert F)$,
$P(F\vert F)$, and $P(N\vert N)$, follow the same logic). When
the results are averaged over all stocks, we obtain
$P(F\vert F)=0.039$ which is much higher than the overall
failure probability $P(F)=0.003$---this is another sign of
volatility clustering in our data. On the level of individual
stocks, however, $62\%$ of all stocks with at least three
failures strongly satisfy the inequality $P(N\vert F)>P(N)$
which is equivalent to $P(F\vert F)<P(F)$ (because
$P(F\vert F)+P(N\vert F)=1$). (By strong satisfying we mean that
the difference of the two probabilities is greater than the sum
of their uncertainties.) We see that despite volatility
clustering in the data, most stocks are more ``resistant'' to
failures after they have just undergone one. For the remaining
stocks, probabilities $P(N\vert F)$ and $P(N)$ either differ
less than the sum of values' uncertainties (for $14\%$ of
stocks) or even strongly satisfy the opposite inequality
$P(N\vert F)<P(N)$, with corresponding values of $P(F\vert F)$
often as high as $0.30$ ($24\%$ of stocks).

To summarize, after a failure (a major price drop), most stocks
become more resistant to another failure---this observation will
serve as a basis for the mathematical model presented in the
following section. At the same time, there is a fraction of
stocks which are prone to consecutive failures---this particular
feature will be discussed in detail in Section~\ref{sec:modif}.

\section{Basic model and its mean-field solution}
In this section we present a basic model which is amenable to
analytical treatment and qualitatively reproduces some of the
features observed in empirical data. In its original
formulation, this model is particularly suitable for stocks
that, as discussed in the previous section, after a failure
become more robust. A generalization of the model aiming at
reproducing other observed features (volatility clustering, for
example) is presented in Section~\ref{sec:modif}.

Consider a system of $N$ nodes where node $i$ ($i=1,\dots,N$)
has only two possible states: failed ($i\in\mathcal{F}$) and
healthy ($i\not\in\mathcal{F}$). With each node $i$ we further
associate fragility $f_i\in[0,1]$ which measures how this node
reacts to failures of its neighbors (the higher the fragility,
the more likely is the node to follow a neighbor's failure). The
dynamics of the model is governed by the following simple rules.
(i) In each time step, the first failed node (``trigger'') is
chosen at  random and may induce failures of other nodes. (ii)
If a neighbor of node $i$ fails, node $i$ follows it with
probability $f_i$ and resists with probability $1-f_i$. (If
several neighbors of node $i$ fail simultaneously, in order to
stay healthy, node $i$ has to resist each individual failure.)
The cascade of failures propagates until all remaining nodes
resist the damage. (iii) At the end of the time step,
fragilities of all nodes are updated according to
\begin{equation}
\label{update}
f_i(t+1)=\Big\{
\begin{aligned}
 \lambda f_i(t) & \qquad i\in\mathcal{F}\\
(1+\beta)f_i(t) & \qquad i\not\in\mathcal{F}
\end{aligned}
\end{equation}
where $0<\beta\ll1$ and $\lambda\in(0,1)$ are parameters of the
model (in effect, failed nodes become less fragile and healthy
nodes become slightly more fragile in the next time step). All
values $f_i(t+1)>1$ are truncated to $1$ (this may occur when
$\beta$ is large). After this update is finished, all nodes are
again marked as healthy, the current time step ends and a new
one begins with point (i). Note that unlike some other models of
cascade spreading, failed nodes are not removed from the system
in our case. If a~long enough equilibration period is applied
before measuring the system behavior, the initial fragility
values $f_i(0)$ are of little importance (see
Section~\ref{sec:initial} for a detailed discussion). Unless
stated otherwise, we set them randomly in the range $(0,1)$ in
our simulations.

According to the rules above, when $n$ neighbors of node $i$
fail, node $i$ resists with the probability $(1-f_i)^n$ and
fails with the complementary probability
\begin{equation}
\label{F-rule}
P_F(f_i,n)=1-(1-f_i)^n.
\end{equation}
This response to failures is ``path-independent'' in some sense:
the probability that a node resists $n$ failures of its
neighbors, $(1-f_i)^n$, is the same as the probability of
resisting two consequent waves of failures of $x$ and $n-x$
neighboring nodes, $(1-f_i)^x(1-f_i)^{n-x}$.

We simplify the system by assuming that interactions of all
nodes are equally strong (the general case will be studied
in Section~\ref{sec:gen}). This renders the notion of ``node's
neighbors'' superfluous because every failure affects all
remaining healthy nodes in the system. Now assume that after the
initial failed node is chosen, $n_1$ nodes respond to this
failure and fail too. Each of the remaining $N-n_0-n_1$ nodes
(here $n_0=1$ is the initial number of failed nodes) then has
some $n_1$-dependent failure probability which results in $n_2$
new failures, and so on, until in iteration $m$, $n_m=0$ is
achieved. The cascade size is then defined as the total number
of failures, $S=n_0+\dots+n_m$, and node fragilities are
consequently updated according to Eq.~\eqref{update}. Since
in one turn nodes can only fail once, cascade sizes are limited
by the system size and $S\leq N$.

The dynamics of the system, based on failure propagation and
fragility updating, is fully contained in the three
above-described rules. In the following paragraphs we shall
study when these rules drive the system to a critical state and
what is the distribution of cascade sizes $P(S)$.

\subsection{Failure probability}
\label{sec:f_prob}
\begin{figure}
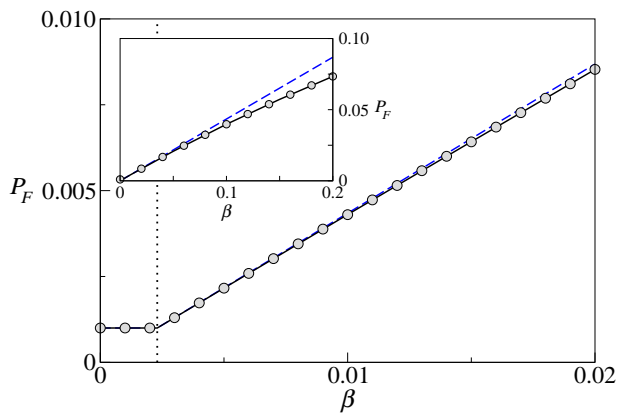

\centering
\vspace*{4pt}
\onefig{P_F-vs-beta}
\caption{Average failure probability: $P_F$ given by
Eq.~\eqref{P_F} (solid black line),
$P_F\approx-\beta/\ln\lambda$ (dashed blue line) and numerical
results (symbols, averaged over $10^6$ time steps) for $N=10^3$,
$\lambda=0.1$. The vertical dotted line indicates $\beta_0$
given by Eq.~\eqref{beta_0}.}
\label{fig:P_B}
\end{figure}

Let $P_F$ be the average failure probability of a given node in
one time step (or, equivalently, the average fraction of failed
nodes in one time step). Assuming that $t_{\mathrm{eq}}$ is
some sufficiently long equilibration time (we use
$t_{\mathrm{eq}}=10^4$ for all our simulations), later fragility
values averaged over realizations, $\avg{f_i}$, do not evolve
anymore. All nodes interact equally strongly, hence $\avg{f_i}$
is independent of $i$ and it can be replaced  with $\avg{f}$.
Since in a~large number of time steps $T$ each node undergoes
$P_FT$ failures and $(1-P_F)T$ non-failures, Eq.~\eqref{update}
implies
\begin{equation}
\label{longT}
\avg{f(t_{\mathrm{eq}}+T)}=\avg{f(t_{\mathrm{eq}})}
\lambda^{P_FT}(1+\beta)^{(1-P_F)T}.
\end{equation}
Using the equilibrium condition
$\avg{f(t_{\mathrm{eq}}+T)}=\avg{f(t_{\mathrm{eq}})}$, we can
solve this equation with respect to $P_F$ to get
\begin{equation}
\label{P_F}
P_F(\beta,\lambda)=
-\frac{\ln(1+\beta)}{\ln\frac{\lambda}{1+\beta}}.
\end{equation}
When $\beta\ll1$, this can be approximated with
$P_F(\beta,\lambda)\approx-\beta/\ln\lambda$ (Fig.~\ref{fig:P_B}
compares these results with numerical simulations).

A node may fail because it is selected as the first failed node
(with probability $1/N$) or due to failure propagation (with
probability $P_P$); $P_F$ thus can be written as $P_F=1/N+P_P$.
Since the value of $P_F$ depends solely on $\beta$ and
$\lambda$, $P_P=P_F-1/N$ may be negative for a small system
which is, of course, impossible in practice. This situation
occurs when for given $\lambda,N$, the value of $\beta$ is
smaller than a certain threshold $\beta_0$ and hence it does not
suffice to compensate for the fragility decay due to $\lambda$.
Eq.~\eqref{longT} then has only the trivial solution $\avg{f}=0$
and hence $P_F(\beta,\lambda)=1/N$ (failures do not spread).
When $\beta$ is small, the approximate form of $P_F$ can be used
to solve this equation with respect to $\beta$ and we get
\begin{equation}
\label{beta_0}
\beta_0\approx-\frac{\ln\lambda}N
\end{equation}
which agrees with numerical simulations (see the vertical line
in Fig.~\ref{fig:P_B}). Note that if the number of initial
failed nodes is assumed to grow with the system size as $wN$
($w\ll1$), we get $\beta_0\approx -w\ln\lambda$ which is
independent of $N$.

When model parameters are set to extreme values (for example,
$N=10^3$, $\beta=10^3$, $\lambda=10^{-3}$), the system exhibits
unusual modes of behavior where active turns (with nearly all
nodes failed) alternate with calm turns (with nearly all nodes
healthy). While Eq.~\eqref{P_F} holds also in such conditions,
our further analysis focuses on $\beta\ll1$ which renders more
realistic behavior.

\subsection{Average fragility}
When $nf_i\ll1$, $P_F$ given by Eq.~\eqref{F-rule} can be
approximated as $P_F(f_i,n)\approx nf_i$ which can be
interpreted as independence of stress inflicted by $n$
individual failed nodes. This further means that each failed
node has its failing descendants independently of other failed
nodes and hence one can use the theory of branching
processes~\cite{Feller70} to describe the cascade spreading.
Note that by use of this theory we implicitly assume that the
system size is infinite. For a discussion of the finite-size
effects on the size of an epidemic outbreak see~\cite{NaKr04}.

As already mentioned, when interactions of all nodes are equal,
$\avg{f_i}$ is independent of $i$. If we further neglect
fluctuations of $f_i$, then all nodes have identical fragility
$\avg{f}$. This is a mean-field-like approximation which
replaces the exact cascade spreading with cascade spreading in a
homogeneous averaged medium. Since the number of direct
descendants now follows a simple binomial distribution with mean
$N\avg{f}$, we can use elementary results of branching process
theory to express the average cascade size (the total progeny)
as $\avg{S}=1/(1-N\avg{f})$. Further, using
$\avg{S}=NP_F(\beta,\lambda)$ we obtain the average fragility
\begin{equation}
\label{f-avg}
\avg{f}=\frac1N\bigg(
1-\frac{\ln\big[(1+\beta)/\lambda\big]}{N\ln(1+\beta)}\bigg).
\end{equation}
Since $\beta>0$ and $\lambda<1$, $\avg{f}$ is always less than
$1/N$. Comparison with numerical simulations (not shown)
confirms that Eq.~\eqref{f-avg} is valid only for $\beta\ll1$.

\subsection{Cascade size distribution}
The theory of branching processes is well studied~\cite{Har89}
and can be easily applied to our model. According to a theorem
from~\cite{Dwass69}, if the generating function for the number
of direct descendants $d$ is $\pi(x)$, the total progeny of the
resulting branching process $Y$ has the distribution
\begin{equation}
P(Y|n_0)=\frac{n_0}Y\,p_{Y-n_0}^{(Y)}
\end{equation}
where $p_a^{(b)}$ is defined using
\begin{equation}
[\pi(x)]^b=p_0^{(b)}+p_{1}^{(b)}x+\dots
\end{equation}
and $n_0$ is the number of ancestors (in our case, the number of
initial failed nodes). Since $d$ obeys a binomial distribution,
its generating function is $\pi(x)=(1-\avg{f}+\avg{f}x)^N$ and
we get
\begin{equation}
\label{P_S}
P(S\vert\beta,\lambda)=
\frac1S\binom{NS}{S-1}\avg{f}^{S-1}
\big(1-\avg{f}\big)^{NS-S+1}
\end{equation}
where we used $n_0=1$ and $\avg{f}$ is given by
Eq.~\eqref{f-avg}. Note that the resulting probability is
positive for $S>N$ which contradicts the model assumptions (each
node fails at most once in a given turn). This is a direct
consequence of using the theory of branching processes which
assumes that the system size is infinite. This problem is of
little importance for small values of $\beta$ when the obtained
values of $P(S)$ are negligible for $S>N$.

\begin{figure}
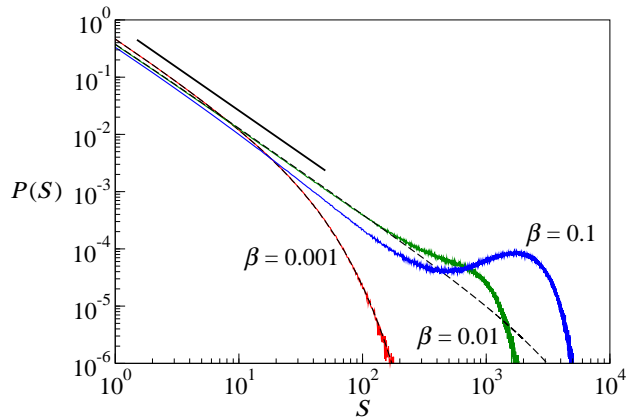

\centering
\vspace*{4pt}
\onefig{P_S}
\caption{The cascade size distribution: numerical results (color
lines), analytical results according to Eq.~\eqref{P_S} (dashed
lines) and the power-law decay with exponent $-3/2$ (thick solid
line) for $N=10^4$, $\lambda=0.1$, $10^7$ time steps, and
$\beta=0.001$ (red line, fastest decay), $\beta=0.01$ (green
line, medium decay), $\beta=0.1$ (blue line, slowest decay). The
analytical solution is not plotted for $\beta=0.1$ because it is
very similar to that for $\beta=0.01$.}
\label{fig:P_S}
\end{figure}

When $1\ll S\ll N$, Eq.~\eqref{P_S} can be approximated with
\begin{equation}
\label{P_S-approx}
P(S\vert\beta,\lambda)=\frac{(N\avg{f})^{S-1}
\mathrm{e}^{S(1-N\avg{f})}}{\sqrt{2\pi}S^{3/2}}.
\end{equation}
According to Eq.~\eqref{f-avg}, $\lim_{N\to\infty}N\avg{f}=1$
for any given $\beta,\lambda$ and hence in the limit of large
system size is $P(S\vert\beta,\lambda)\sim S^{-3/2}$ which
corresponds to the classical critical branching process. For a
finite system, the smaller the value of $\beta$, the larger the
value of $1-N\avg{f}$. Consequently, the power-law scaling holds
only for $S\ll\beta N$ (this agrees with Fig.~\ref{fig:P_S}
where for $\beta=10^{-3}$, the power-law behavior disappears at
$S\approx10$). On the other hand, the range of $\beta$ and
$\lambda$ for which the system self-organizes to a critical
state is wide and we can say that this is an SOC system.

A comparison of the obtained analytical results with numerical
simulations is shown in Fig.~\ref{fig:P_S}. The agreement is
good for small values of $\beta$ ($\beta\lesssim0.01$) and the
initial slope of the distributions (before the finite-size
effects become apparent) is close to $-3/2$. Results obtained
with $\beta=0.001$ confirm that when $\beta$ is small enough,
$P(S)$ decays faster than as a power law. When $\beta$ is large,
true $P(S)$ deviates from the analytical prediction and exhibits
a secondary maximum at a large size value---this effect is well
visible in Fig.~\ref{fig:P_S} for $\beta=0.1$. This maximum,
formally simply a super-critical phase of the model, resembles
so-called meaningful outliers discussed in~\cite{Sorn09}. To
estimate the value of $\beta$ at which the secondary maximum
appears and Eq.~\eqref{P_S} ceases to hold, we take the average
number of failures computed both from Eq.~\eqref{P_S} and
from Eq.~\eqref{P_F}. By comparing the two results we obtain
\begin{equation}
\label{avg_S-comparison}
NP_F(\beta,\lambda)=\sum_{S=1}^{N} SP(S\vert\beta,\lambda).
\end{equation}
When $\beta$ is small, both sides of this equation depend on
$\beta$ and the equality can hold. However,
Eq.~\eqref{P_S-approx} shows that when $\beta$ is sufficiently
large, the size distribution is approximately power-law and it
is independent of $\beta$. As we increase $\beta$ further, the
power-law distribution does not suffice to provide enough
failures and for Eq.~\eqref{avg_S-comparison} to hold, an
additional contribution must appear on the rights side. The
value $\beta_1$ when this happens can be found by substituting 
$P(S)\sim S^{-3/2}$ on the right side and approximating the
summation with integration. When $N$ is large, we obtain
\begin{equation}
\label{beta_star}
\beta_1\approx-\bigg(\frac2{\pi N}\bigg)^{1/2}\ln\lambda
\end{equation}
which complements the previously found threshold $\beta_0$. For
$N=10^4$ and $\lambda=0.1$, we obtain $\beta_1\approx0.02$ which
agrees with our empirical observation ($\beta\lesssim0.01$ for
Eq.~\eqref{P_S} to hold) above.

Finally, by comparing the empirical observations presented in
Fig.~\ref{fig:real_stocks} with the obtained analytical results,
we can conclude that the presented model exhibits qualitative
agreement with the studied real system.

\subsection{Generalizations}
\label{sec:gen}
To test how robust are the obtained results, we consider simple
generalizations of the proposed model. First of all, when the
multiplicative fragility update rule Eq.~\eqref{update} is
replaced by an additive one, the behavior of the system does not
change considerably. The second generalization relates to the
assumed even influence of a node's failure on all the remaining
nodes. Denoting the strength of failure propagation from node
$i$ to node $j$ as $C_{i,j}$, the probability that node $j$
fails as a result of $i$'s failure can be generalized to
$C_{i,j}f_j$. The probability that node $j$ fails as a result of
a~group $\mathcal{F}$ of failed nodes (given by
Eq.~\eqref{F-rule} before) generalizes to the form
\begin{equation}
\label{F-ruleC}
P_F(f_j,\mathcal{F})=1-\prod_{i\in\mathcal{F}}(1-C_{i,j}f_j).
\end{equation}
Matrix $\mathsf{C}$ encodes the structure of the network of node
interactions.

When the elements $C_{i,j}$ are drawn independently from a~given
distribution and the system size is large, the mean-field
approximation is again appropriate to describe the system
behavior and the power-law size distribution with exponent $3/2$
results. Similarly when $\mathsf{C}$ contains a block structure
with inter-block elements drawn from a different distribution
than intra-block elements (this mimics the sector structure of
the stock correlation matrix~\cite{ManSta99,KimJe05}), the
original power-law size distribution remains largely unchanged
(unless either the block division of $\mathsf{C}$ or one of the
two probabilistic distributions are such that they do not allow
to use the mean-field approximation). Analogous behavior results
from the ``random neighbor approximation'' in which node's
neighbors are chosen anew repeatedly (see~\cite{BDFJW94} for
this kind of analysis of a~different model).

\begin{figure}
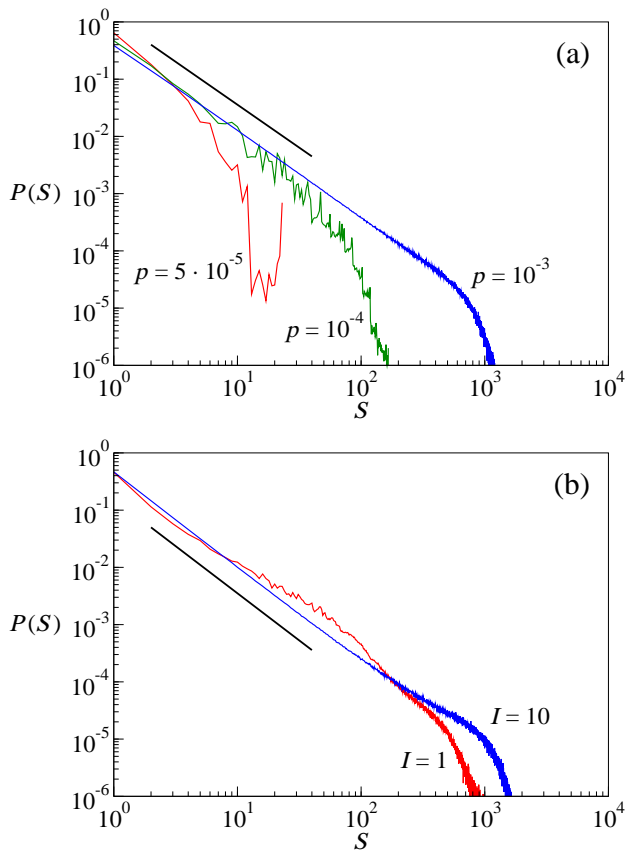

\centering
\vspace*{4pt}
\onefig{P_S-erd}\\[6pt]
\onefig{P_S-bara}
\caption{The cascade size distribution on complex networks: (a)
sparse Erdős-Rényi networks with
$p=5\cdot10^{-5},10^{-4},10^{-3}$; the indicative thick line has
slope $1.5$, (b) Barabási-Albert networks with $I=1$ and $I=10$;
the indicative thick line has slope $1.65$). Parameters of the
system: $N=10^4$, $\beta=0.005$, $\lambda=0.1$, $10^7$ time
steps.}
\label{fig:networks}
\end{figure}

When all elements $C_{i,j}$ are either zero or one, matrix
$\mathsf{C}$ can be represented by a network and a complex
topology of node interactions can be introduced by network
models~\cite{Newman03}. We studied two different types of
networks: the Erdős-Rényi network where $C_{i,j}=1$ with
probability $p$ and $C_{i,j}=0$ otherwise and the growing
Barabási-Albert network where each new node is attached to $I$
old nodes. (These two kinds of networks are structurally very
distinct as the former consists of nodes of approximately
identical degree and the latter exhibits a power-law degree
distribution.) Numerical results for both cases are shown in
Fig.~\ref{fig:networks}. As expected, for the Erdős-Rényi
network with $p>1/N$, the size distribution exponent remains
unchanged. When $p<1/N$, the network consists of small isolated
components and hence big cascades cannot occur. The irregular
size distribution $P(S)$ observed for $\beta=5\cdot10^{-5}$ is
due to topological properties of the particular network
realization where the model was simulated (\emph{i.e.},
positions of respective ups and downs of the size distribution
depend on the network realization). These results agree with a
previous study of the sandpile dynamics~\cite{Bona95}
(see~\cite{DoGoMe08} for an extensive recent review of critical
phenomena in complex networks). By contrast, Barabási-Albert
networks yield cascade size distributions with significantly
higher exponents (approximately $1.65$) which is probably due to
strong inhomogeneity of the network. When $I=1$, $P(S)$ deviates
from a power law, probably as a~consequence of the scale-free
network topology (the same shape of the distribution is observed
for different realizations of the network).

\subsection{Role of the initial fragility values}
\label{sec:initial}
\begin{figure}
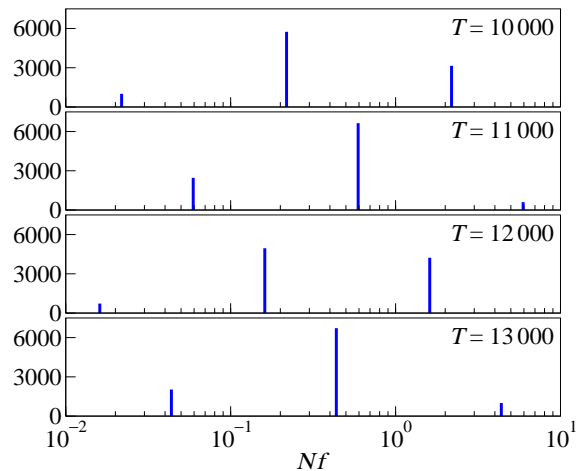

\centering
\vspace*{4pt}
\onefig{f_profiles}
\caption{Fragility distributions at different time steps (the
initial fragility values are set to $1/N$, $N=10^4$,
$\beta=0.001$, $\lambda=0.1$).}
\label{fig:profiles}
\end{figure}

While it sounds plausible that due to model's stochasticity, the
initial fragility values have no influence on the equilibrium
fragility distribution, the situation is in fact more
complicated. For example, a simple numerical simulation with
$f_i(0)=1/N$ for all $i$ shows a case where: (i) no stationary
fragility distribution arises, (ii) at any time step, only a
small number of distinct fragility values is observed (see
Fig.~\ref{fig:profiles}). What causes the discreteness of
fragility values? Denoting the number of failing and healthy
time steps of node $i$ as $F_i$ and $H_i$, respectively, it must
hold that $F_i+H_i=t$ where $t$ is the current time step. This
node's fragility now can be written as
\begin{equation}
\label{f-evol}
f_i(t)=f_i(0)(1+\beta)^t\big[\lambda/(1+\beta)\big]^{F_i}. 
\end{equation}
When all $f_i(0)$ are identical, the possible values of $f_i(t)$
are discrete at any time step $t$ and the ratio of neighboring
possible values is $(1+\beta)/\lambda$. If $\lambda$ is small
(as it is in our simulations), this ratio is large and hence the
number of actually observed fragility values is small (because
values much smaller or greater than the average fragility are
unlikely). Eq.~\eqref{f-evol} implies that possible fragility
values depend on $t$ and hence there can be no stationary
fragility distribution---this is confirmed by
Fig.~\ref{fig:profiles} where fragility peaks constantly shift
to higher values and change their relative heights.
Interestingly, even this peculiar setting of $f_i(0)$ does not
alter the long-term model's behavior substantially and the
aggregate quantities (such as the average failure probability or
the cascade size distribution) are similar to those found for
randomized initial fragility values before.

Differences between neighboring peaks are $\lambda/(1+\beta)$,
hence the time after which the fragility distribution pattern
repeats can be estimated as
$\ln\big[\lambda/(1+\beta)\big]/\ln(1+\beta)$. Since this is a
typical time of fragility evolution, one can use it also as an
estimate of the initial equilibration time $T_{\mathrm{eq}}$.
For the smallest value of $\beta$ in our simulations
($\beta=0.005$) we obtain $T_{\mathrm{eq}}\approx 4\,600$ which
ex post confirms our setting of the equilibration time to
$10^4$. Finally, note that while the random setting of $f_i(0)$
prevents discrete fragility values from appearing, some remnants
of the initial fragility values can be preserved by
Eq.~\eqref{f-evol}. To obtain a fragility distribution truly
independent of the initial values, one has to assume annealed
dynamics, \emph{i.e.} fragility updating by randomized values of
$\beta$ and $\lambda$.

\section{Generalized model with partial memory}
\label{sec:modif}
Fragility updating rules defined by Eq.~\eqref{update} imply
that nodes become more robust after a failure and hence
autocorrelation of their failures as well as autocorrelation of
the total number of failures are negative (their magnitudes
depend on $\beta$ and $\lambda$). As discussed in
Section~\ref{sec:empirical}, this is true for majority of stocks
but certainly not for all of them. To allow for repeatedly
failing stocks, we introduce the probability $\alpha$ with which
a failed node stays failed also in the next time step (and
consequently acts as an additional initial failed node). This
probability has the role of \emph{partial memory} in the system 
and, as we shall see, gives rise to volatility clustering and
other effects observed in real financial data. Note that memory
or delayed stress propagation are quite often part of cascade
spreading models as in, for example, \cite{PeBuHe08}. We assume
that fragilities of nodes which stay failed due to $\alpha$ are
not updated in the given time step (when $\alpha$ is small, this
assumption has little influence on the results).

\begin{figure}
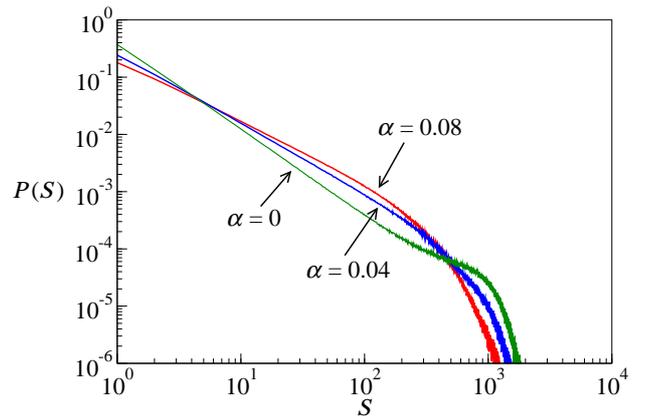

\centering
\vspace*{4pt}
\onefig{P_S2}
\caption{Cascade size distribution for the modified model:
numerical results for $\beta=0.01$, $\lambda=0.1$, $N=10^4$,
$10^7$ time steps, and various values of $\alpha$.}
\label{fig:P_S-gen}
\end{figure}

Since $P(F)\ll1$, the probability of a node's repeated failure
is now $P(F\vert F)\approx\alpha$ which, in combination with the
empirical results presented in Section~\ref{sec:empirical},
motivates us to set $\alpha=0.04$. We further choose
$\beta=0.01$ and $\lambda=0.1$ which best correspond to the
critical regime in Fig.~\ref{fig:P_S}. Using this setting we
numerically obtain conditional probabilities consistent with
those observed in the empirical data: $P(F\vert F)=0.041$
(empirical value is $0.039$), $P(F)=0.004$ (empirical value is
$0.003$) and $P(F\vert N)=0.004$ (empirical value is $0.003$).
As long as we stay in the critical regime, these values depend
on $\beta$ and $\lambda$ weakly. Presence of volatility
clustering is confirmed by significantly positive
autocorrelation of the number of failures
$C(n_F(t),n_F(t+1))\approx0.3$ (empirical value is $0.15$). The
precise value depends on $\alpha$ and $\lambda$ (and much less
on $\beta$) but positive autocorrelation naturally arises for
$\alpha$ which is large enough. By contrast, $\alpha=0$ yields
$P(F\vert F)\approx P(F)$ and $C(n_F(t),n_F(t+1))\approx-0.04$.
Finally, Fig.~\ref{fig:P_S-gen} shows $P(S)$ for different
values of $\alpha$. We see that for small values of $\alpha$,
the size distribution remains power-law with exponent gradually
decreasing as $\alpha$ grows. Due to the additional complexity
introduced by partial memory, an analytical cascade size
distribution for this generalized model has not been obtained
yet.

\section{Discussion}
We studied empirical stock prices and found that large
simultaneous downturns follow a broad distribution consistent
with a power law with exponent $2.19\pm0.05$. To reproduce this
behavior, we proposed a minimal stochastic model of failure
propagation. Using a mean field approximation and branching
process theory we derived the general cascade size distribution
and determined the range of parameters which give rises to the
critical regime. To reproduce other features observed in
financial data, such as volatility clustering, partial memory
was introduced within the basic model.

While our model implicitly assumes arrival of news to the market
(they cause the initial nodes to fail and allow cascades to be
created), we minimize the influence of news on the system's
behavior by assuming their equal impact (in each time step,
exactly one initial node is chosen to fail). This approach is
motivated by the extensive study of \emph{excess volatility}
which shows that it is difficult to link the observed trading
volumes and volatility to the arriving information~\cite{MiMu94}
and even the large crash of 1987 does not seem to be triggered
by particular news~\cite{Black88}. In reality, of course, the
impact of news on the market differs from one day to another. It
could be therefore interesting to test how different ways of
choosing the initial failed nodes influence the model's behavior
(for example, the number of the initial nodes can be random or
network hubs may be preferentially chosen to trigger a cascade).

There is a number of other challenging questions which deserve
further investigation. Firstly, since the cascade sizes
corresponding to the secondary maximum in Fig.~\ref{fig:P_S} are
comparable with system size, this behavior cannot be described
within the formalism of branching processes where an infinite
system size is assumed. While we found an approximate condition
for the appearance of the secondary maximum, how to proceed
further towards an exhaustive description of the resulting size
distribution is still an open question. Secondly, it would be
interesting to find an analytical expression for the size
distribution exponent in scale-free networks where it appears to
differ from the mean-field value $3/2$. Thirdly, generalized
model with ``partial memory'', studied only numerically here,
calls for analytical approaches. Fourthly, it would be
interesting to know whether the model can be modified to produce
power-law size distributions with exponents considerably higher
than those reported here. One opportunity for such a
generalization is to assume a dynamic network structure whose
evolution depends on nodes' failures, similarly to the approach
used in~\cite{BiMa04,CaGa09} for different models.
Alternatively, as a generalization of the current binary model
where nodes are either healthy or failed, one could define a
multi or continuous-state model in which the probability of
following a neighbor's failure depends on the failure's
magnitude.

We stress that the probabilistic spreading mechanism proposed
here is a general one and its use is not limited to market
crashes or firm bankruptcies. For example, economic exchanges
between countries are so intense that decline in one country may
propagate to a neighboring one (take, for example, how growth in
many European countries depends on spending of German
consumers). On a two or three dimensional lattice, a similar
mechanism might be employed to model earthquakes because,
similarly to the proposed model, a failure at one place of the
Earth's crust exerts some stress on its neighborhood (the number
of failed nodes would then represent the earthquake size). In
summary, the proposed model, together with its generalizations,
has proven to be simple yet rich in behavior. It poses a variety
of new research questions and we are looking forward to its
future development and applications.

\begin{acknowledgement}
This work was partially supported by the Future and Emerging
Technologies programme FP7-COSI-ICT of the European Commission
through project QLectives (grant no. 231200) and by the Swiss
National Science Foundation (project no. 200020-121848). We
acknowledge insightful suggestions of Matteo Marsili, enjoyable
and helpful discussions with Damien Challet and Chi Ho Yeung,
and comments of our anonymous reviewers.
\end{acknowledgement}

\end{document}